\documentclass[12pt, letterpaper]{JHEP3}

\usepackage{amssymb, amsmath, amsopn, amsthm}

\usepackage{epsfig}

\usepackage{epsfig,multicol}

\usepackage{amsmath}
\usepackage{graphicx}
\usepackage{latexsym}
\usepackage{amsfonts}
\usepackage{amssymb}

\newcommand{\eref}[1]{(\ref{#1})}

\newcommand{\fref}[1]{Figure~\ref{#1}}
\newcommand{\cref}[1]{Chapter~\ref{#1}}
\newcommand{\beq}{\begin{equation}}
\newcommand{\eeq}{\end{equation}}
\newcommand{\ba}{\begin{array}}
\newcommand{\ea}{\end{array}}
\newcommand{\bcenter}{\begin{center}}
\newcommand{\ecenter}{\end{center}}

\def\IB{\relax\hbox{$\inbar\kern-.3em{\rm B}$}}
\def\IC{\relax\hbox{$\inbar\kern-.3em{\rm C}$}}
\def\ID{\relax\hbox{$\inbar\kern-.3em{\rm D}$}}
\def\IE{\relax\hbox{$\inbar\kern-.3em{\rm E}$}}
\def\IF{\relax\hbox{$\inbar\kern-.3em{\rm F}$}}
\def\IG{\relax\hbox{$\inbar\kern-.3em{\rm G}$}}
\def\IGa{\relax\hbox{${\rm I}\kern-.18em\Gamma$}}
\def\IH{\relax{\rm I\kern-.18em H}}
\def\IK{\relax{\rm I\kern-.18em K}}
\def\IL{\relax{\rm I\kern-.18em L}}
\def\IP{\relax{\rm I\kern-.18em P}}
\def\IR{\relax{\rm I\kern-.18em R}}
\def\IZ{\relax\ifmmode\mathchoice
{\hbox{\cmss Z\kern-.4em Z}}{\hbox{\cmss Z\kern-.4em Z}}
{\lower.9pt\hbox{\cmsss Z\kern-.4em Z}}
{\lower1.2pt\hbox{\cmsss Z\kern-.4em Z}}\else{\cmss Z\kern-.4em Z}\fi}
\def\II{\relax{\rm I\kern-.18em I}}

\def\sCC{{\kern 0.27em\vrule height1.45ex width0.03em depth0em
          \kern-0.30em\rm C}}
\def\C{{\mathchoice
  {\sCC}
  {\sCC}
  {\kern 0.225em \vrule height1.05ex width0.025em depth0em \kern-0.25em \rm C}
  {\kern 0.180em \vrule height0.78ex width0.02em depth0em \kern-0.2em \rm C}
        }}
\def\sHH{{\rm I\kern-.16em{}H}}
\def\H{{\mathchoice
  {\sHH}
  {\sHH}
  {\rm I\kern-.13em{}H}
  {\rm I\kern-.13em{}H} }}
\def\sNN{{\rm I\kern-.16em{}N}}
\def\N{{\mathchoice
  {\sNN}
  {\sNN}
  {\rm I\kern-.12em{}N}
  {\rm I\kern-.10em{}N} }}
\def\sPP{{\rm I\kern-.16em{}P}}
\def\P{{\mathchoice
  {\sPP}
  {\sPP}
  {\rm I\kern-.12em{}P}
  {\rm I\kern-.10em{}P} }}
\def\sQQ{{\kern 0.27em \vrule height1.45ex width0.03em depth0em
          \kern-0.30em \rm Q}}
\def\Q{{\mathchoice
        {\sQQ}
        {\sQQ}
  {\kern 0.225em \vrule height1.05ex width0.025em depth0em \kern-0.25em \rm Q}
  {\kern 0.180em \vrule height0.78ex width0.020em depth0em \kern-0.20em \rm Q}
        }}
\def\sRR{{\rm I\kern-0.16em{}R}}
\def\R{{\mathchoice
  {\sRR}
  {\sRR}
  {\rm I\kern-0.12em{}R}
  {\rm I\kern-0.10em{}R} }}
\def\sZZ{{\rm Z\kern-0.32em{}Z}}
\def\Z{{\mathchoice
  {\sZZ}
  {\sZZ} 
  {\rm Z\kern-0.3em{}Z}     
  {\rm Z\kern-0.25em{}Z} }}  
\def\ZZZ{{\rm Z\kern-0.24em{}Z}}
\def\sII{{\rm I\kern-0.16em{}I}}
\def\I{{\mathchoice
  {\sII}
  {\sII}
  {\rm I\kern-0.12em{}I}
  {\rm I\kern-0.10em{}I} }}

\def\Tr{{\rm Tr}}

\def\inbar{\,\vrule height1.5ex width.4pt depth0pt}
\font\cmss=cmss10 \font\cmsss=cmss10 at 7pt

\def\smiley{\hbox{\large$\bigcirc$\hspace{-0.80em}\raise.2ex
\hbox{$\cdot\cdot$}\kern-.61em\lower.2ex\hbox{\scriptsize$\smile$}}\ }
\def\frowny{\hbox{\large$\bigcirc$\hspace{-0.80em}\raise.2ex
\hbox{$\cdot\cdot$}\kern-.635em\lower.2ex\hbox{\scriptsize$\frown$}}\ }

\def\I{{\rlap{1} \hskip 1.6pt \hbox{1}}}

\makeatletter
\let\hangafter\@hangfrom
\makeatother

\def\makeatletter{\catcode`\@=11}
\makeatletter
\def\mathbox#1{\hbox{$\m@th#1$}}%
\def\math@ccstyles#1#2#3#4#5#6#7{{\leavevmode
     \setbox0\mathbox{#6#7}%
     \setbox2\mathbox{#4#5}%
     \dimen@ #3%
     \baselineskip\z@\lineskiplimit#1\lineskip\z@
     \vbox{\ialign{##\crcr
            \hfil \kern #2\box2 \hfil\crcr
            \noalign{\kern\dimen@}%
            \hfil\box0\hfil\crcr}}}}
\def\mathaccstyles{\math@ccstyles\maxdimen}
\def\maththroughstyles{\math@ccstyles{-\maxdimen}}
\def\unity%
{\maththroughstyles{.45\ht0}\z@\displaystyle {\mathchar"006C}\displaystyle 1}

\renewcommand{\H}{\mathcal{H}}

\newcommand{\be}{\begin{eqnarray}}
\newcommand{\bea}{\begin{eqnarray}}
\newcommand{\ee}{\end{eqnarray}}
\newcommand{\eea}{\end{eqnarray}}

\newcommand{\bb}{\mathsf{b}}
\newcommand{\ww}{\mathsf{w}}

\title{Dimer Models and Integrable Systems}

\author{Richard Eager$^{1,2}$, Sebasti\'an Franco$^3$ and Kevin Schaeffer$^4$

\\

\vspace{0.6cm}
$^1$Department of Physics,
University of California, Santa Barbara, CA 93106, USA\\
\vspace{0.cm}
$^2$Institute for the Physics and Mathematics of the Universe, University of Tokyo \\
Kashiwa, Chiba 277-8582, Japan \\
\vspace{0.2cm}
$^3$Kavli Institute for Theoretical Physics, University of California \\
Santa Barbara, CA 93106, USA \\
\vspace{0.2cm}
$^4$Center for Theoretical Physics, Department of Physics,
University of California \\ Berkeley, CA 94720, USA \\
\vspace{0.3cm}

\email{richard.eager@ipmu.jp, sebastian.franco@durham.ac.uk}\\

}

\abstract{We explore various aspects of the correspondence between dimer models and integrable systems recently introduced by Goncharov and 
Kenyon. Dimer models give rise to relativistic integrable systems that match those arising from 5d $\mathcal{N}=1$ gauge theories studied by Nekrasov. We apply the correspondence to dimer models associated to the $Y^{p,0}$ geometries, showing that they give rise to the relativistic generalization of the periodic Toda chain originally studied by Ruijsenaars. The correspondence reduces the calculation of all conserved charges to a straightforward combinatorial problem of enumerating non-intersecting paths in the dimer model. We show how the usual periodic Toda chain emerges in the non-relativistic limit and how the Lax operator corresponds to the Kasteleyn matrix of the dimer model. We discuss how the dimer models for general $Y^{p,q}$ manifolds give rise to other 
relativistic integrable systems, generalizing the periodic Toda chain and construct the integrable systems for general $Y^{p,p}$ explicitly. The impurities introduced in the construction of $Y^{p,q}$ quivers are identified with impurities in twisted $\mathfrak{sl}(2)$ XXZ spin chains. Finally we discuss how the physical concept of higgsing a dimer model provides an efficient method for producing new integrable systems starting from known ones. We illustrate this idea by constructing the integrable systems for higgsings of $Y^{4,0}$.}

\preprint{IPMU 11-0112} 

\begin{document}

\tableofcontents

\section{Introduction}

Integrable systems have a dense web of connections with gauge theories and string theory. Integrable systems appear in a variety of contexts including Seiberg-Witten theory \cite{Gorsky:1995zq}, recent connections to the vacua of supersymmetric theories \cite{Nekrasov:2009rc}, and the calculation of the spectrum of anomalous dimensions \cite{Gromov:2009bc} and scattering amplitudes \cite{Alday:2010vh} in super-Yang-Mills, to name a few.  

Recently, Goncharov and Kenyon discovered an exciting correspondence between integrable systems and dimer models \cite{GK}. According to their correspondence, every dimer model defines an integrable system, whose conserved charges can be systematically calculated from perfect matchings.

The correspondence sheds new light on integrable systems, with applications and implications yet to be investigated. It provides new perspectives on integrable systems that are naturally related to dimer models, such as 4d quiver gauge theories, D3-branes probing toric Calabi-Yau 3-folds \cite{Franco:2005rj}, mirror symmetry \cite{Feng:2005gw} and quantum Teichm\"uller space \cite{GK,Franco:2011sz}.

In this paper, we investigate various aspects of the correspondence from a physical perspective. This work is organized as follows. Section 2 briefly reviews the relation between dimer models and quiver theories, toric Calabi-Yaus and integrable systems. Section 3 discusses the connection between integrable systems and 5d $\mathcal{N}=1$ and 4d $\mathcal{N}=2$ gauge theories. In Section 4, we apply the correspondence, constructing a relativistic generalization of the periodic Toda chain from the dimer models associated to $Y^{p,0}$ manifolds. We study the relation between the Kasteleyn and Lax operators, the non-relativistic limit of the integrable system and how dimer models for general $Y^{p,q}$ geometries produce alternative relativistic generalizations of the periodic Toda chain. We also discuss the connection to twisted $\mathfrak{sl}(2)$ XXZ spin chains with impurities. In Section 5, we introduce a practical method for generating new integrable systems based in higgsing and illustrate the method with explicit examples. We conclude and mention future directions in Section 6.

\bigskip

\section{Some Background}

\label{section_background}

In this section we provide a lightning review of various concepts used throughout this paper. When necessary, we indicate references for more thorough explanations.

\subsection*{Dimers and Quivers}

Brane tilings, to which we will also refer to as dimer models, are bipartite graphs embedded in a two-torus. The dual of a brane tiling is a planar, periodic quiver. There is a one-to-one correspondence between brane tilings and periodic quivers \cite{Franco:2005rj} that is summarized in the following dictionary:

\medskip

\beq
\begin{array}{ccccc}
\mbox{{\bf Gauge Theory}} & & & & \mbox{{\bf Brane Tiling}} \\
\mbox{gauge group} & \ \ \ \ & \leftrightarrow & \ \ \ \ & \mbox{face}\\
\mbox{chiral superfield} & \ \ \ \ & \leftrightarrow & \ \ \ \ & \mbox{edge} \\
\mbox{superpotential term} & \ \ \ \ & \leftrightarrow & \ \ \ \ & \mbox{node}
\end{array}
\nonumber
\eeq

\bigskip

\begin{figure}[h]
\begin{center}
\includegraphics[width=12cm]{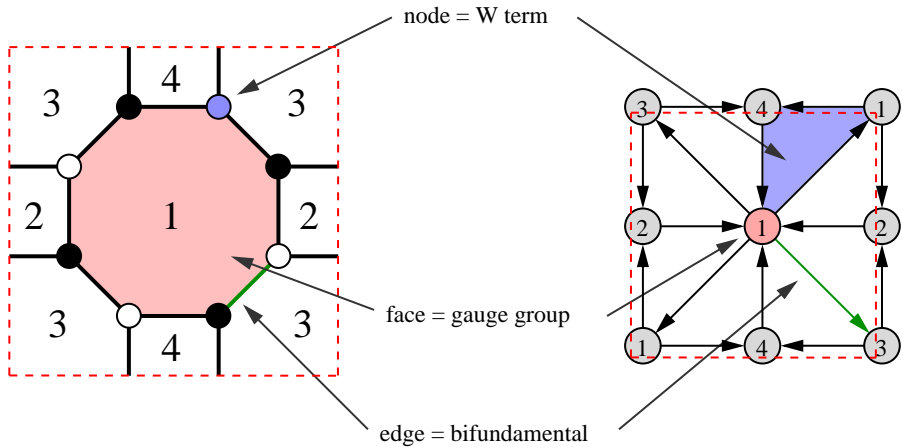}
\caption{Correspondence between the brane tiling and the periodic quiver for phase I of $F_0$.}
\label{tiling_quiver_F0}
\end{center}
\end{figure}

Every term in the superpotential of the gauge theory is encoded in an oriented plaquette of the periodic quiver. \fref{tiling_quiver_F0} exemplifies the correspondence for phase I of $F_0$ \cite{Feng:2002zw}.

\subsection*{Dimers and Calabi-Yau 3-folds}

Quiver gauge theories that are described by brane tilings arise on the worldvolume of stacks of D3-branes probing singular, toric Calabi-Yau (CY) 3-folds. The CY geometry emerges as the moduli space of vacua of the quiver gauge theory. The connection between dimer models and quivers has trivialized the determination of the corresponding CY geometry. GLSM fields in the toric description of the CY are in one-to-one correspondence with perfect matchings of the dimer model. As a result, points in the toric diagram correspond to (sets of) perfect matchings. This correspondence reduces the task of finding the CY geometry to computing the determinant of the Kasteleyn matrix \cite{Franco:2005rj}. 

\bigskip

\subsection*{Dimers and Integrable Systems}

The dynamical variables of the integrable system correspond to oriented loops in the brane tiling. One basis for such loops is given by the cycles going clockwise around each face $w_i$ ($i=1,\ldots,N_g$, with $N_g$ the number of gauge groups in the quiver) and the cycles $z_1$ and $z_2$ wrapping the two directions of the 2-torus.\footnote{Since $\prod_{i=1}^{N_g} w_i=1$, one of the $w_i$'s is redundant. This identity can also be exploited for simplifying expressions.}$^,$\footnote{The analysis of some models, such as the ones in Section \ref{section_periodic_Toda}, can be considerably simplified by choosing a different basis.}
 
The Poisson brackets between basis cycles are
\beq
\begin{array}{ccl}
\{w_i,w_j\} & = & \epsilon_{w_i,w_j} \, w_i w_j \\
\{z_1,z_2\} & = & (\langle z_1,z_2 \rangle +\epsilon_{z_1,z_2}) \, z_1 z_2 \\
\{z_a,w_i\} & = & \epsilon_{z_a,w_i} \, z_a w_i
\end{array}
\label{Poisson_brackets}
\eeq
where $\epsilon_{x,y}$ is the number of edges on which the loops $x$ and $y$ overlap, counted with orientation. Then, $\epsilon_{w_i,w_j}$ is simply the antisymmetric oriented incidence matrix of the quiver. In addition, $\langle z_1,z_2 \rangle$ is the intersection number in homology of the cycles $z_1$ and $z_2$.

The classical integrable system can be quantized replacing the Poisson brackets by a q-deformed algebra of the form
\beq
X_i X_j = q^{\{x_i,x_j\}} X_j X_i \, ,
\label{commutators}
\eeq
where $X_i=e^{x_i}$ and $q=e^{-i2\pi \hbar}$.

Every perfect matching corresponds to a point in the toric diagram and defines a closed loop by subtraction of a reference perfect matching. This loop can then be expressed in terms of the basic cycles. When multiple perfect matchings correspond to a given toric diagram point, their contributions must be added. Goncharov and Kenyon showed that the commutators defined by \eref{commutators} and \eref{Poisson_brackets} result in a $(0+1)$-dimensional quantum integrable system in which the conserved charges are given by:

\bigskip

\begin{itemize}
\item{\bf Casimirs:} they commute with everything. They are defined as the ratio between contributions of consecutive points on the boundary of the toric diagram.

\item{\bf Hamiltonians:} they commute with each other and correspond to internal points in the toric diagram. 
\end{itemize}

\bigskip

In this paper we will not discuss the choice of reference perfect matching in detail, we instead refer the interested reader to \cite{GK}. Different choices of the reference perfect matching correspond to shifts in the toric diagram. These overall shifts do not affect Casimirs (since they are defined as ratios of points in the toric diagram) but they modify the Hamiltonian(s). Models with zero or one internal point are insensitive to this choice, since they have zero or one Hamiltonians.  The choice of reference perfect matching becomes important for models with more than one internal point and can be straightforwardly determined by demanding the Hamiltonians to commute.

Following \cite{Kenyon:2003uj} (see also \cite{Franco:2006gc} for applications), we define the magnetic flux through a loop, $\gamma$, in terms of edges in the tiling as

\beq
v(\gamma)=\prod_{i=1}^{k-1} {X(\ww_i,\bb_i)\over X(\ww_{i+1},\bb_i)} \, ,
\label{flux_v}
\eeq
where the product runs over the contour $\gamma$ and $\bb_i$ and $\ww_j$ denote black and white nodes. With this definition, the elements in our basis are $w_j \equiv v(\gamma_{w_j})$, $z_1\equiv v(\gamma_{z_1})$, and $z_2\equiv v(\gamma_{z_2})$. Another brief but more complete summary of the work in \cite{GK} can be found in \cite{Franco:2011sz}.

\section{Integrable Systems from Dimers, 5d and 4d}

\subsection*{5d Gauge Theories and Dimers}

The integrable systems we are discussing can be derived from either dimer models or 5d gauge theories. The main object underlying all constructions is the spectral curve $\Sigma$. 

Dimer models encode the quiver gauge theory (and also the geometry) on D3-branes probing a singular, toric Calabi-Yau 3-fold X. The toric singularity has a characteristic polynomial $P(z_1,z_2)=\sum a_{n1,n2} \, z_1^{n_1}z_2^{n_2}$, where $(n_1, n_2)$ runs over points in the toric diagram. The mirror Calabi-Yau is given by $P(z_1,z_2)=W$, $W= u v$. In this case, $\Sigma$ corresponds to the Riemann surface sitting at $W=0$ in the mirror.

Nekrasov \cite{Nekrasov:1996cz} proposed a non-perturbative solution for 5d gauge theories compactified on a circle using relativistic integrable systems.  From the Seiberg-Witten solution \cite{Seiberg:1994rs,Seiberg:1994aj} a connection to integrable systems was proposed in \cite{Gorsky:1995zq}. In particular, the spectral curve of the integrable system matches the Seiberg-Witten curve.  Nekrasov's insight was to generalize the integrable system to a relativistic integrable system in order to determine the corresponding Seiberg-Witten curve for the 5d gauge theory.  
There are two possible ways to engineer such compactified 5d theories and $\Sigma$ plays a prominent role in both of them.  First, we can construct them by wrapping an M5-brane on $\Sigma$. This M5-brane is a de-singularization of a web of $(p,q)$ 5-branes in Type IIB, obtained after compactifying in a circle of radius $\beta$ to pass to Type IIA and then lifting to M-theory \cite{Aharony:1997bh}. Alternatively, these theories can be obtained in the low energy limit of M-theory on a Calabi-Yau 3-fold with vanishing cycles to decouple gravity \cite{Morrison:1996xf,Douglas:1996xp,Ganor:1996pc,Intriligator:1997pq}. The Calabi-Yau is the same X of the dimer construction, so we see that $\Sigma$ is also relevant from this perspective. Particle-like states arise from M2-branes wrapped around SUSY 2-cycles.

\subsection*{4d Gauge Theories and the Non-Relativistic Limit}

The integrable systems constructed using dimer models (equivalently 5d gauge theories compactified on a circle) are naturally {\it relativistic}. This is reflected, for example, in the exponential dependence on momenta of the conserved charges. This fact is manifest in the form of Poisson brackets \eref{Poisson_brackets}.

\begin{figure}[h]
\begin{center}
\includegraphics[width=14cm]{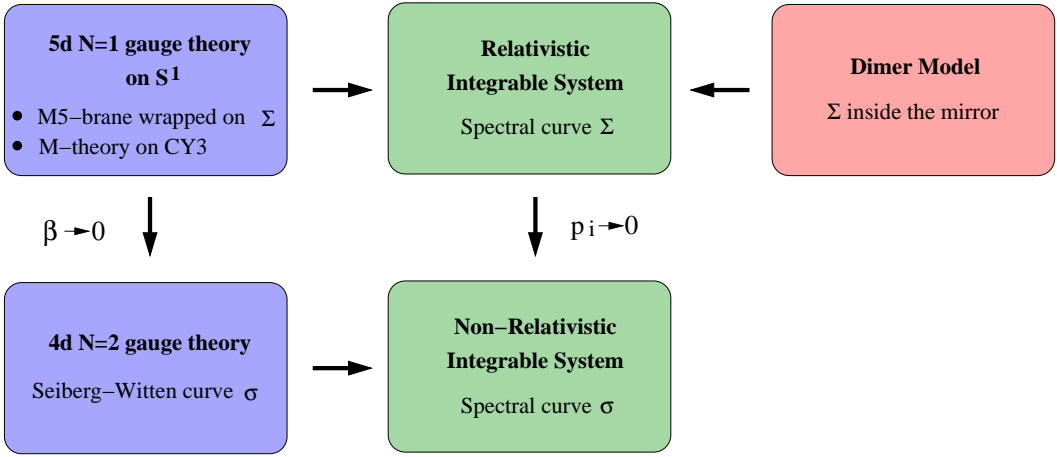}
\caption{Dimer models and 5d, $\mathcal{N}=1$ gauge theories compactified on an $S^1$ correspond to different perspectives on the spectral curve $\Sigma$. The 4d, $\mathcal{N}=2$ limit of the gauge theory corresponds to a non-relativistic limit of the integrable system.}
\label{5d-dimer}
\end{center}
\end{figure}

The radius, $\beta$, of the circle on which the 5d gauge theory is compactified plays the role of the inverse speed of light. The non-relativistic limit corresponds to taking $\beta \to 0$.\footnote{Also interesting is the decompactification limit $\beta \rightarrow \infty$, in which the 5d perturbative solution \cite{Antoniadis:1995vz} is recovered} In this limit, we obtain a 4d, $\mathcal{N}=2$ gauge theory whose Seiberg-Witten curve is the spectral curve of the non-relativistic integrable system, which we denote $\sigma$.\footnote{As explained in Section \ref{section_background}, dimer models are in one-to-one correspondence with 4d (generically $\mathcal{N}=1$) quiver theories. These quivers thus provide other 4d gauge theories naturally associated to the integrable systems. In what follows, whether we refer to the 4d $\mathcal{N}=2$ or quiver gauge theories should be clear from the context.} \fref{5d-dimer} summarizes the connection between 5d gauge theories, dimer models, integrable systems and their non-relativistic (4d) limits.

\medskip

\section{The Periodic Toda Chain}

\label{section_periodic_Toda}

In this section we show the correspondence from \cite{GK} at work in explicit examples, using dimer models to construct (relativistic generalizations of) the periodic Toda chain. The relativistic, periodic Toda chain was first introduced in \cite{Ruijsenaars} and studied in connection with 5d gauge theories in 
\cite{Nekrasov:1996cz}. These integrable systems arise from dimer models, equivalently quiver gauge theories, associated to $Y^{p,q}$ manifolds \cite{Benvenuti:2004dy}. The associated spectral curves, $\Sigma$, correspond to 5d, $\mathcal{N}=1$, $SU(p)$ gauge theory with no flavors and different values of a quantized parameter controlling the cubic couplings in the prepotential.

\subsection{$Y^{p,0}$ Integrable Systems}

\label{section_Yp0}

\subsection*{Integrable system}

\begin{figure}[h]
\begin{center}
\includegraphics[width=4cm]{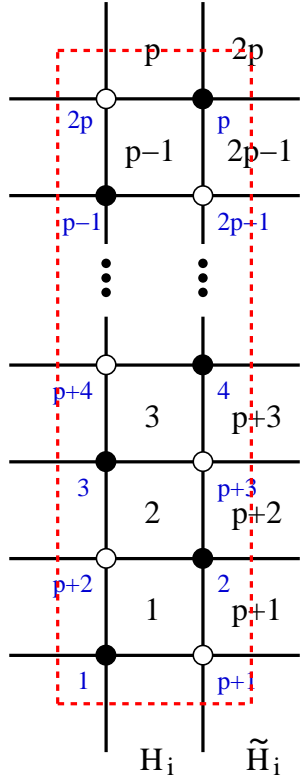}
\caption{Brane tiling for $Y^{p,0}$ with even $p$. We indicate the numbering of nodes in blue.}
\label{dimer_Yn0}
\end{center}
\end{figure}

Let us now consider general $Y^{p,0}$ geometries. Our goal is to make contact with the relativistic Toda chain for arbitrary $p$. For concreteness, let us focus on the case in which $p$ is even. The shape of the unit cell depends on whether $p$ is even (rectangle) or odd (rhombus). The dimer model for the conifold corresponds to a square lattice \cite{Hanany:2005ve}. The cone over $Y^{p,0}$ is a $\mathbb{Z}_p$ orbifold of the conifold. As a result, its dimer model is given by a square lattice with an enlarged unit cell, given by $p$ copies of the one for the conifold, as shown in \fref{dimer_Yn0}.\footnote{A similar configuration for a discrete time integrable system related to dimer models was considered in \cite{Korepanov:1994rc}.}

\begin{figure}[h]
\begin{center}
\includegraphics[width=9cm]{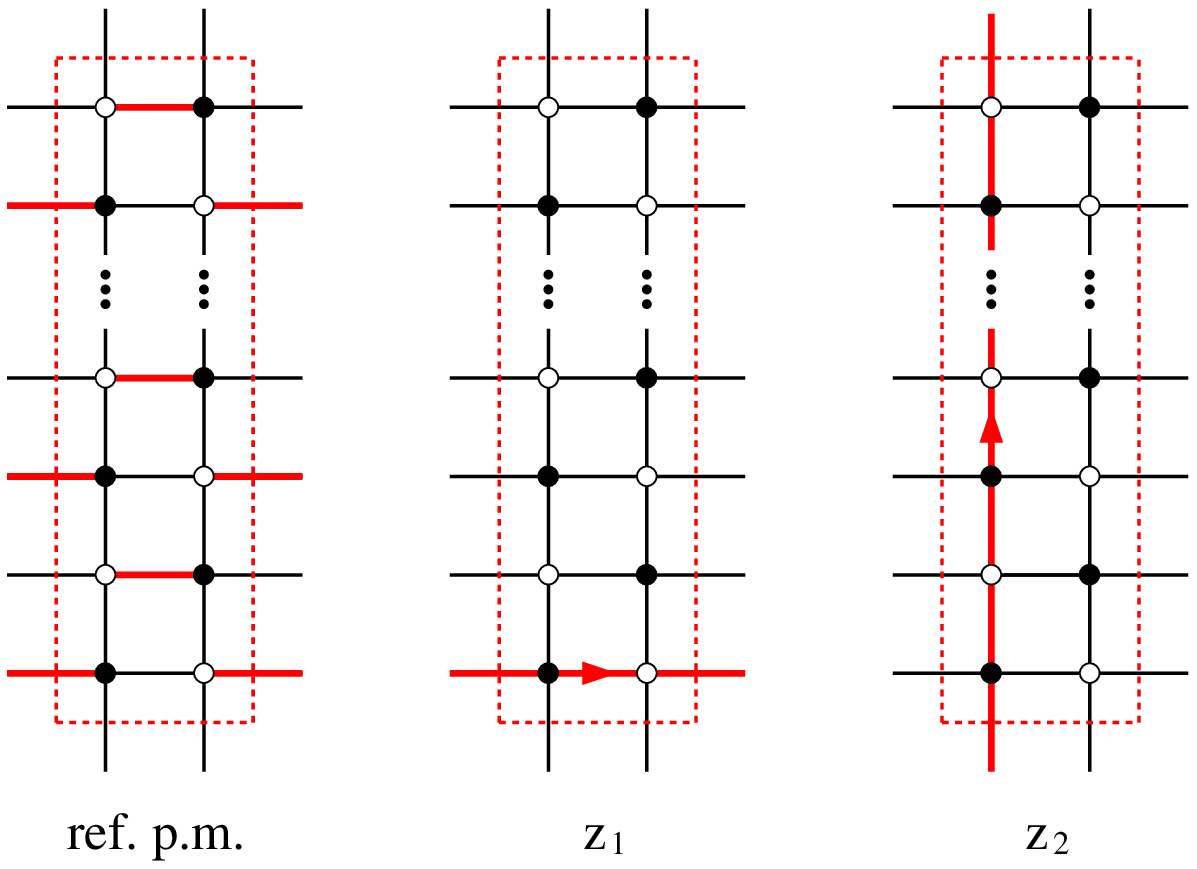}
\caption{Reference perfect matching and $z_1$ and $z_2$ paths for $Y^{p,0}$.}
\label{reference_z1_z2_Yn0}
\end{center}
\end{figure}

\begin{figure}[h]
\begin{center}
\includegraphics[width=6cm]{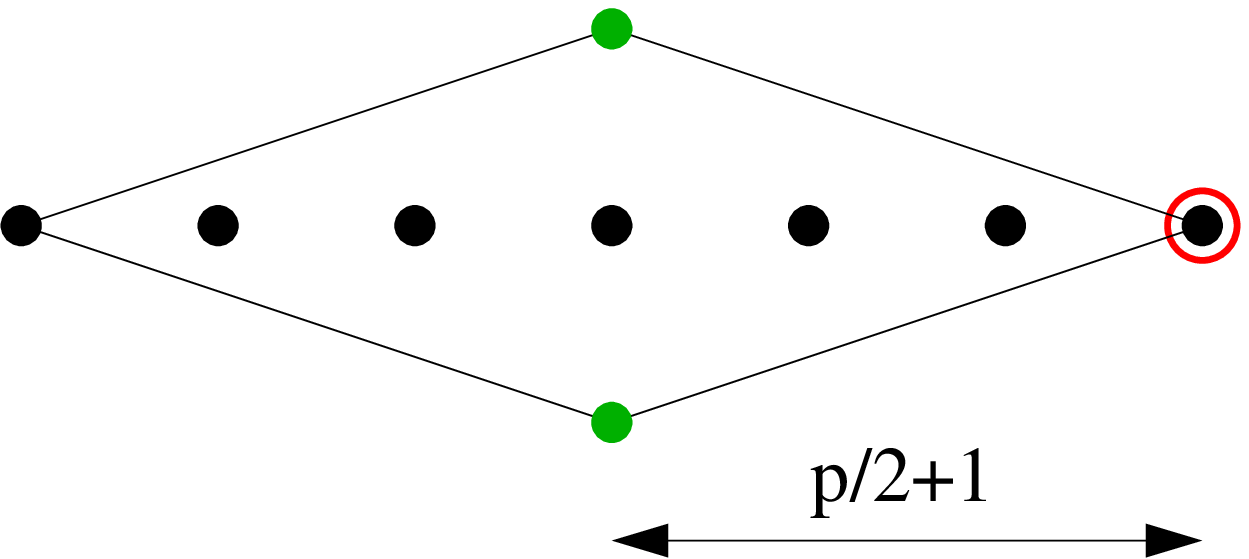}
\caption{Toric diagram for $Y^{p,0}$ (shown in the figure for $p=6$). The reference perfect matching is circled in red. By construction, its position in the $(z_1,z_2)$ plane is $(0,0)$. The green dots correspond to cycles with windings $(-p/2-1,1)$ and $(-p/2-1,-1)$.}
\label{toric_general_Yp0}
\end{center}
\end{figure}

The reference perfect matching and $z_1$ and $z_2$ paths for $Y^{p,0}$ are shown in \fref{reference_z1_z2_Yn0}. \fref{toric_general_Yp0} shows the toric diagram for $Y^{p,0}$, with even $p$. The system has a $\mathbb{Z}_2$ symmetry corresponding to choosing the opposite corner of the toric diagram as the reference perfect matching. The $\mathbb{Z}_2$ symmetry interchanges the Hamiltonian with its dual \cite{Kharchev:2001rs}.

\begin{figure}[h]
\begin{center}
\includegraphics[width=9.5cm]{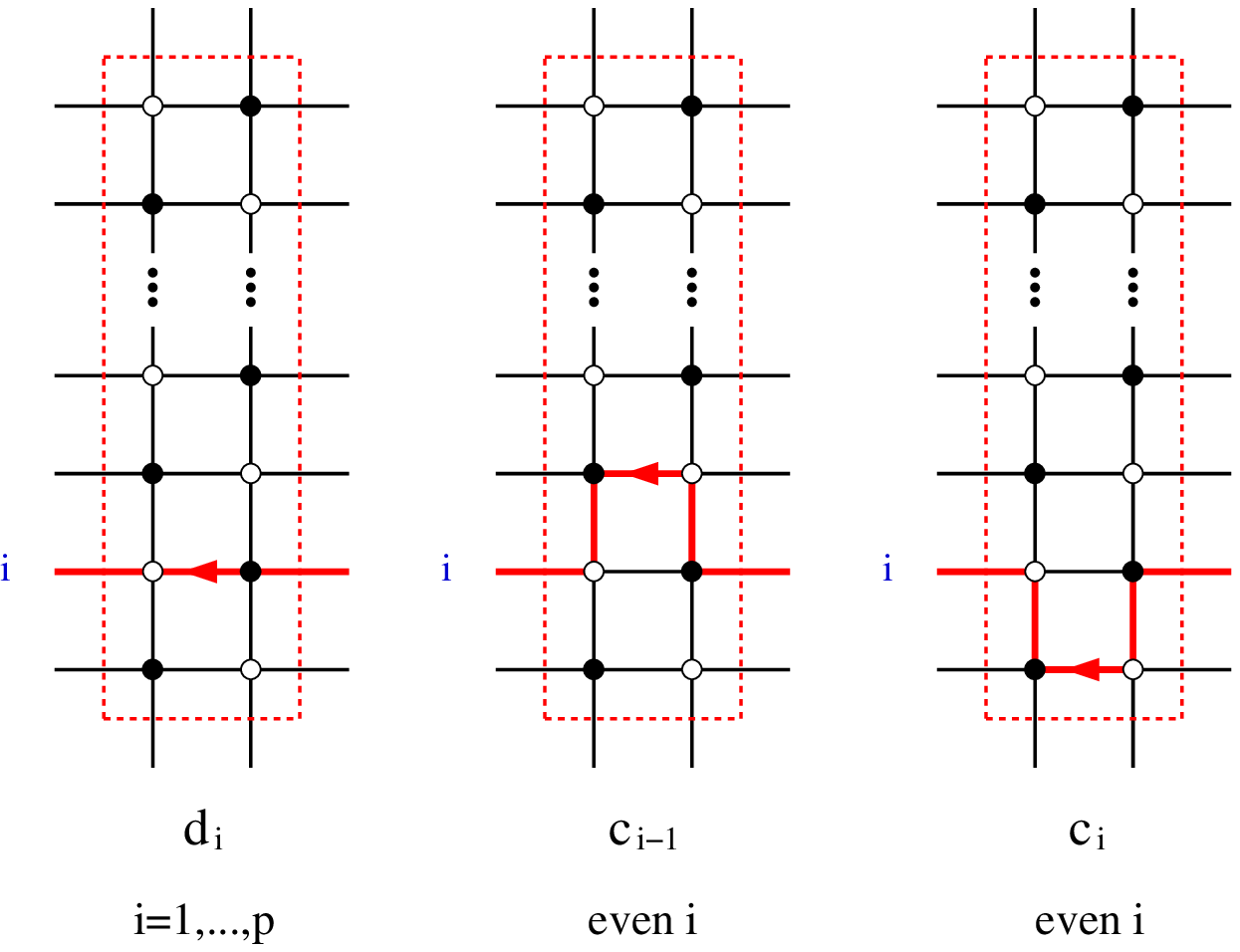}
\caption{$2p$ of the cycles in a convenient basis for $Y^{p,0}$. Notice that the $c$ cycles only exist for even $i$.}
\label{cycles_Yn0}
\end{center}
\end{figure}

The construction of the associated integrable system is considerably simplified by an appropriate choice of basis of $2p+2$ cycles, instead of $w_i$ and $z_j$. \fref{cycles_Yn0} shows $2p$ of the loops. There are two additional cycles, with winding numbers $(-p/2-1,1)$ and $(-p/2-1,-1)$ around the $z_1$ and $z_2$ directions, which correspond to the green points in \fref{toric_general_Yp0}. They are mapped to two Casimirs and are hence fixed.

The non-vanishing Poisson brackets are

\beq
\{c_k,d_k\}=c_k d_k \ \ \ \ \ \ \ \ \{c_k,d_{k+1}\}=-c_k d_{k+1} \ \ \ \ \ \ \ \ \{c_k,c_{k+1}\}=-c_k c_{k+1}
\label{PB_Yp0}
\eeq
The Hamiltonian corresponds to the $(-1,0)$ point in the toric diagram and is given by
\beq
H_1=\sum_{i=1}^{p} (c_i+d_i) \,.
\label{H_Yp0}
\eeq
Equations \eref{PB_Yp0} and \eref{H_Yp0} precisely agree with the Poisson brackets and Hamiltonian for the general periodic relativistic Toda chain \cite{Bruschi:1988xp,Kharchev:2001rs}. The $c_i$ and $d_i$ variables can be expressed in terms of position and momentum variables with canonical commutation relations as follows

\begin{eqnarray}
c_i & = & \exp(p_i-q_i+q_{i+1}) \nonumber \\
d_i & = & \exp p_i \, .
\label{qp_1}
\end{eqnarray}

Using the $c_i$ and $d_i$ cycles, determining the additional $(p-2)$ higher Hamiltonians reduces to a straightforward combinatorial problem. The $H_n$ Hamiltonian, associated with the $(-n,0)$ point in the toric diagram, corresponds to the sum of all possible combinations of $n$ of these cycles with the condition that they do not overlap or touch at any vertex of the tiling. For example, for $Y^{4,0}$ we have:

\begin{eqnarray}
H_2 & = & \left(c_1 c_3 + c_2 c_4 \right) + \left(c_1(d_3 + d_4) + c_2(d_4 + d_1) + c_3(d_1 + d_2) + c_4(d_2 + d_3)  \right) \\
 & + & \left(d_1 d_2 + d_1 d_3 + d_1 d_4 + d_2 d_3 + d_3 d_4 \right)\, , \nonumber \\ \nonumber \\
H_3 & = & (d_2 d_3 d_4  + d_1 d_3 d_4  + d_1 d_2 d_4 + d_1 d_2 d_3) + (c_1 d_3 d_4 + c_2 d_4 d_1 + c_3 d_1 d_2 + c_4 d_2 d_3) \, . \nonumber
\end{eqnarray}

\bigskip

\subsection{Kasteleyn matrix}

\label{section_Kasteleyn_Toda}

Let us construct the Kasteleyn matrix for the dimer model in \fref{dimer_Yn0}. It is convenient to label edges according to whether they are horizontal ($H$ and $\tilde{H}$) or vertical ($V$ and $\tilde{V}$). $H$ edges are those at the center of the tiling and $\Tilde{H}$ are those crossing the edge of the unit cell. In addition:

\beq
\begin{array}{cl}
V & \mbox{: vertical with black node at top endpoint} \\
\tilde{V} & \mbox{: vertical with white node at top endpoint} 
\end{array}
\nonumber
\eeq

Finally, let us call $H_i$ and $\tilde{H}_i$ the edges connecting nodes $i$ and $n+i$, $V_i \equiv V_{i+1,p+i}$ and $ \tilde{V}_i \equiv \tilde{V}_{i,p+i+1}$. Subindices indicate nodes on the tiling and are identified ${\rm mod}(2p)$. In these variables, the Kasteleyn matrix is

\beq
K_p=\left(\begin{array}{c|cccccc}
 & p+1 & p+2 & \ \ \ \ p+3 \ \ \ \ & \cdots & \ \ \ 2p-1 \ \ \ & 2p\\ \hline 
1 & -H_1-\tilde{H}_1 \, z_1 & \tilde{V}_1 &  &  & & V_p \, z_2 \\
2 & V_1 & H_2+\tilde{H}_2 \, z_1^{-1} & \tilde{V}_2 & & &  \\ 
3 & & V_2 & \ddots & &  &  \\ 
\vdots & & & & \ddots \\
p-1 & & & & & \ddots & \tilde{V}_{p-1} \\
p & \tilde{V}_p \, z_2^{-1} & & & & V_{p-1} & H_p + \tilde{H}_p \, z_1^{-1} 
\end{array}\right)
\label{K_Yn0}
\eeq
\bigskip
Notice the alternating overall sign of the terms on the diagonal.


\subsubsection{Non-Relativistic Limit}

The Kasteleyn matrix \eref{K_Yn0} and the Lax operator of the non-relativistic periodic Toda chain \cite{Flaschka1,Flaschka2} are strikingly similar. We now make this connection explicit by first expressing the edge variables in terms of coordinates and momenta and then taking the non-relativistic (i.e. small momentum) limit.\footnote{The definitions in \eref{VH_qp} are reasonable, rather symmetric and simple. We later check that they are indeed consistent with the $w_i$ commutation relations.} We define

\beq
\begin{array}{ccccc}
V_i & = & \tilde{V}_i & \equiv & e^{q_i-q_{i+1}}\\
H_i & = & -\tilde{H_i}^{-1} & \equiv & e^{(-1)^{i} p_i/ 2}
\end{array}
\label{VH_qp}
\eeq
and
\beq
z_1 \equiv e^{-z} \ \ \ \ \ \ z_2 \equiv w
\eeq
In the next section, we explain how these definitions and canonical Poisson brackets $\{p_i,q_j\}=\delta_{ij}$ are in agreement with the Goncharov-Kenyon Poisson brackets.\footnote{Changing from the basis of coordinates and momenta in \eref{VH_qp} to the one in \eref{qp_1} is straightforward.}

Taking the small momenta limit (i.e. linear order in $p_i$ and $z$) of the Kasteleyn matrix \eref{K_Yn0}, we conclude that

\beq
\det K_p = \det (L_p(w)-z) 
\eeq
with
\beq
L_p(w)=\left(\begin{array}{cccccc}
 p_1 & e^{q_1-q_2} &  &  & &  e^{q_p-q_1}\, w \\
e^{q_1-q_2} & p_2 & e^{q_2-q_3} & & &  \\ 
 & e^{q_2-q_3} & \ddots & &  &  \\ 
 & & & \ddots \\
 & & & & \ddots & e^{q_{p-1}-q_p} \\
e^{q_p-q_1} \, w^{-1} & & & & e^{q_{p-1}-q_p} & p_p
\end{array}\right)
\label{Lax_Yp0}
\eeq
which is precisely the Lax operator of the non-relativistic periodic Toda chain. 
Notice that the previous analysis nicely associates coordinates and momenta with the vertical and horizontal directions of the square lattice, respectively.

\subsubsection{Loop Poisson Brackets from Edges}

In the previous section we have expressed edges of the brane tiling in term of coordinates and momenta in such a way that the Lax operator of the periodic Toda chain is obtained from the Kasteleyn matrix of the dimer model by taking the non-relativistic limit. We now show how the commutation relations among loop variables given by the Goncharov-Kenyon rules are recovered from our edge definitions and the $\{p_i,q_j\}$ Poisson brackets.

Consider $Y^{p,0}$ with even $p$. Using equation \eref{flux_v}, we have

\beq
\begin{array}{lccllllcccllllc}
\mbox{odd i:} \ \ \  & w_i  & =  & H_i & V_i^{-1} & H_{i+1} & \tilde{V}_i^{-1} & \ \ \ \ \ \ \ \ \ & 
w_{p+i}  & = & \tilde{H}_i^{-1} & \tilde{V}_i &  \tilde{H}_{i+1}^{-1} & V_i & \\
\mbox{even i:} & w_i  & =  & H_i^{-1} & V_i & H_{i+1}^{-1} & \tilde{V}_i & \ \ \ \ \ \ \ \ \ & 
w_{p+i}  & = & \tilde{H}_i & \tilde{V}_i^{-1} & \tilde{H}_{i+1} & V_i^{-1} & 
\end{array}
\label{ws_Yn0_pq}
\eeq
$i=1,\ldots,p$. From \eref{ws_Yn0_pq} and the canonical Poisson brackets $\{p_i,q_j\}=\delta_{ij}$ (which become $[p_i,q_j]=-i \hbar \, \delta_{ij}$ in the quantum theory) we can calculate

{\footnotesize
\beq
{\{A,B\}\over A B}=\left(\begin{array}{c|cccccc|cccccc} 
& \ w_1 \ & \ w_2 \ & \ \cdots \ & \ \cdots \ & \ \cdots \ & \ w_p \ &  \ w_{p+1} \ & \ w_{p+2} \ & \ \cdots \ & \ \cdots \ & \ \cdots \ & \ w_{2p} \ \\ \hline
\ w_1 \ &  & 2 & & & & 2 & -4 & & & & & \\
\ w_2 \ & -2 &  & -2 & & & & & 4 & & & & \\
\ w_3 \ & & 2 & & 2 & & & & & -4 & & & \\
\ \vdots \ & & & \ddots & & \ddots & & & & & \ddots & & \\
\ w_{p-1} \ & & & & 2 & & 2 & & & & & -4 & \\
\ w_p \ & -2 & & & & -2 & & & & & & & 4 \\ \hline
\ w_{p+1} \ & 4 & & & & & & & -2 & & & & -2 \\
\ w_{p+2} \  & & -4 & & & & & 2 &  & 2 & & &\\
\ w_{p+3} \  & & & 4 & & & & & -2 & & -2 & & \\
\ \vdots \ & & & & \ddots & & & & & \ddots & & \ddots & \\
\ w_{2p-1} \ & & & & & 4 & & & & & -2 & & -2 \\
\ w_{2p} \ & & & & & & -4 &2 & & & & 2 &
\end{array}\right)
\label{w_PB_from_pq}
\eeq}
\bigskip

\noindent which, modulo an unimportant overall scaling, are exactly the Poisson brackets that follow from Goncharov-Kenyon prescription! This scaling can be absorbed in the value of $\hbar$ in the quantum theory. It is important to notice that the final result depends crucially on the details of \eref{VH_qp}, such as the sign of the exponents, which are also vital for obtaining the correct non-relativistic limit. An interesting example that depends on these details is $\{w_i,w_{p+i+1}\}=0$.\footnote{Of course, the variables in \eref{VH_qp} also reproduce the Poisson brackets \eref{PB_Yp0} for the basis of cycles we considered in Section \ref{section_Yp0}. Those cycles are typically written in terms of coordinates and momenta such that horizontal lines are equal to $e^{p'_i}$ and squares are equal to $e^{q'_i-q'_{i+1}}$. The $p_i$ and $q_i$ considered in this section are the ones in which the Lax operator takes the form \eref{Lax_Yp0}, but can be mapped to $p_i'$ and $q_i'$ by an appropriate change of variables.}

\subsection{More Relativistic Generalizations of Toda from $Y^{p,q}$}

\label{section_integrable_from_Ypq}

\begin{figure}[h]
\begin{center}
\includegraphics[width=4cm]{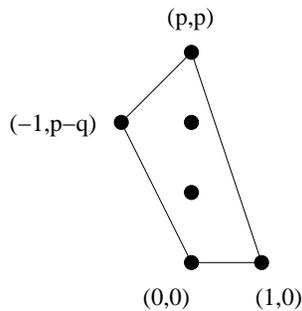}
\caption{Toric diagram for the cone over $Y^{p,q}$.}
\label{toric_Ypq}
\end{center}
\end{figure}

The integrable models based on $Y^{p,0}$ are not the only possible relativistic generalizations of the periodic Toda chain. In fact, as we now discuss, 
all $Y^{p,q}$ geometries give rise to valid generalizations. The corresponding toric diagram is shown \fref{toric_Ypq}.\footnote{The toric diagram in \fref{toric_general_Yp0} can be put into this form by an $SL(2,\mathbb{Z})$ transformation.} These integrable systems were conjectured to exist in \cite{Brini:2008rh}.

M-theory on the CY cones over $Y^{p,q}$ geometries gives rise to 5d, $\mathcal{N}=1$, pure $SU(p)$ gauge theories. The theories differ in the value of $c_{cl}$, which is a quantized parameter of the theory that controls the cubic couplings in the exact quantum prepotential, related to a five dimensional Chern-Simons term \cite{Intriligator:1997pq}. For $Y^{p,q}$, we have $c_{cl}=q$.

The distinction between theories with different values of $c_{cl}$ disappears when taking the non-relativistic limit. This fact is clearly manifest at the level of the spectral curve. The spectral curve associated with $Y^{p,q}$ and the manipulations for taking its non-relativistic limit have repeatedly appeared in the literature (see for example \cite{Hollowood:2003cv}, which we now follow).\footnote{In fact, this correspondence of the spectral curve is the original reason why we identify $Y^{p,q}$ 
dimer models as giving rise to relativistic generalizations of the periodic Toda chain.} The spectral curve, $\Sigma$, can be written as

\beq
z_1+\alpha {z_2^{p-q}\over z_1}+P_p(z_2)=0
\eeq
with $P_p(z_2)$ a degree $p$ polynomial. We can rewrite $\Sigma$ as

\beq
y^2=\prod_{i=1}^p (z_2 - e^{\phi_i})^2-4 e^{-t_B} z_2^{p-q}
\eeq
In order to take the non-relativistic (4d) limit, we define 
\beq
z_2=e^{\beta x} \ , \ \ \ \ \ \ \ e^{\phi_i}=e^{\beta a_{i,i+1}} \ , \ \ \ \ \ \ \ e^{-t_B}=\left({\beta \Lambda \over 2} \right)^{2p} \ ,
\eeq
with $a_{i,i+1}=a_i-a_{i+1}$. From our perspective, it is clear that not only $z_2$ but also the $\phi_i$ variables and $t_B$ must be rescaled when taking the non-relativistic limit since they are controlled by the $w_i$ variables, which are momentum dependent. In the $\beta\to 0$ limit, $\Sigma$ becomes
\beq
y^2=\prod_{i=1}^p (x - a_{i,i+1})^2-4\left({\Lambda\over 2}\right)^{2p} \, ,
\eeq
which is the Seiberg-Witten curve for the pure $\mathcal{N}=2$ $SU(N)$ gauge theory. As we have anticipated, the dependence on $q$ has disappeared. Reversing the reasoning, we conclude that all $Y^{p,q}$ manifolds (i.e. for arbitrary values of $q$) give rise to integrable systems that can be considered relativistic generalizations of the periodic Toda chain. In the next section we discuss the $Y^{p,p}$ geometry and its corresponding integrable system.

\subsection{$Y^{p,p}$ Integrable Systems}

\label{section_Ypp}

We now construct a new infinite family of relativistic integrable systems associated with $Y^{p,p}$. From the discussion in section \ref{section_integrable_from_Ypq}, we know that these systems also reduce to the periodic Toda chain in the non-relativistic limit. 

\begin{figure}[h]
\begin{center}
\includegraphics[width=9.5cm]{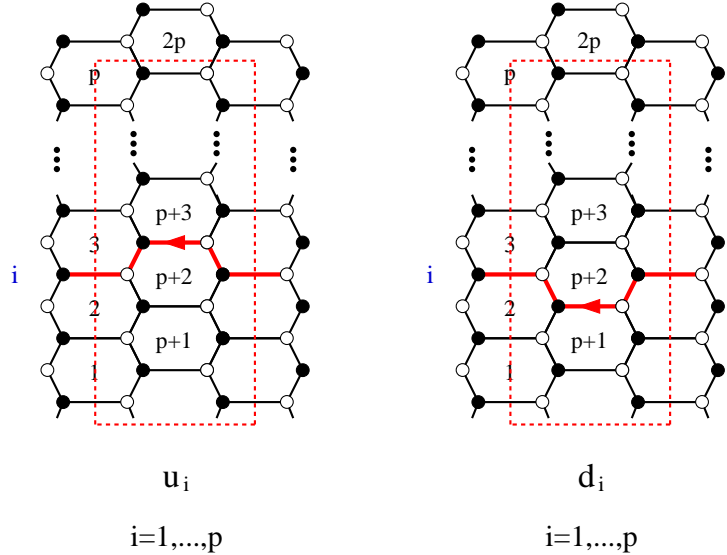}
\caption{Brane tiling for $Y^{p,p}$. A convenient basis  for $Y^{p,p}$ is given by the $u_i$ (up) and $d_i$ (down) cycles, $i=1,\ldots,p$.}
\label{cycles_Ypp}
\end{center}
\end{figure}

The cone over $Y^{p,p}$ is the $\mathbb{C}^3/\mathbb{Z}_{2p}$ orbifold. Its brane tiling consist of two columns of $p$ hexagons \cite{Franco:2005rj} as shown in \fref{cycles_Ypp}.

\begin{figure}[h]
\begin{center}
\includegraphics[width=3cm]{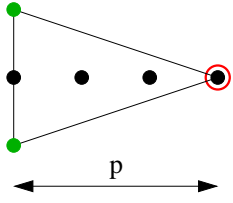}
\caption{Toric diagram for $Y^{p,p}$ (shown in the figure, $p=3$). The reference perfect matching is circled in red. By construction, its position in the $(z_1,z_2)$ plane is $(0,0)$. The green dots correspond to cycles with windings $(-p,1)$ and $(-p,-1)$.}
\label{toric_general_Ypp}
\end{center}
\end{figure}

\fref{toric_general_Ypp} shows the toric diagram for $Y^{p,p}$, where the reference perfect matching is indicated with a red circle. As for $Y^{p,0}$, the analysis of these models is simplified by a convenient choice of basis for closed cycles. \fref{cycles_Ypp} shows $2p$ of them. There are two additional cycles with windings $(-p,1)$ and $(-p,-1)$ along the $(z_1,z_2)$ directions. They correspond to the green points in \fref{toric_general_Ypp} and are fixed by the Casimirs.

The Hamiltonian associated to the $(-1,0)$ point in the toric diagram takes the simple form

\beq
H_1=\sum_{i=1}^p (u_i+d_i) \, .
\eeq
Similar to our analysis of $Y^{p,0}$, determining the Hamiltonians for the $(-n,0)$ points is very simple in the $u_i$ and $d_i$ basis. Finding the $n^{th}$ Hamiltonian reduces to determining all possible combinations of $n$ cycles that do not overlap or intersect at nodes of the tiling. For example, the higher Hamiltonians for $Y^{4,4}$ are

\begin{eqnarray}
H_2 & = & u_1 u_2 +u_1 u_3 +u_1 u_4 +u_2 u_3 +u_2 u_4 +u_3 u_4 \nonumber \\
    & + & d_1 d_2 +d_1 d_3 +d_1 d_4 +d_2 d_3 +d_2 d_4 +d_3 d_4 \nonumber \\
    & + & u_1 d_3 +u_1 d_4 +u_2 d_4 +u_2 d_1 +u_3 d_1 +u_3 d_2 +u_4 d_2+u_4 d_3 \, , \nonumber \\
\nonumber \\
H_3 & = & u_1 u_2 u_3 + u_1 u_2 u_4 +u_1 u_3 u_4 +u_2 u_3 u_4  \nonumber \\
    & + & u_1 d_3 d_4 + u_2 d_4 d_1 +u_3 d_1 d_2 +u_4 d_2 d_3  \nonumber \\
    & + & d_1 d_2 d_3 + d_1 d_2 d_4 +d_1 d_3 d_4 +d_2 d_3 d_4  \nonumber \\
    & + & d_1 u_2 u_3 + d_2 u_3 u_4 +d_3 u_4 u_1 +d_4 u_1 u_2  \, , \nonumber \\
\nonumber \\
H_4 & = & u_1 u_2 u_3 u_4 + d_1 d_2 d_3 d_4 \, .
\end{eqnarray}

\subsection{$Y^{p,q}$ Integrable Systems as Spin Chains}

We have identified the family of integrable systems associated to the $Y^{p,q}$ dimer models in the previous section. Previously the integrable systems identified with the $Y^{p,q}$ spectral curves were described as twisted $\mathfrak{sl}(2)$ XXZ spin chains with impurities \cite{Gorsky:1997mw}.  We will now explain the equivalence of these apparently different descriptions.

The $\mathfrak{sl}(2)$ spin chains are described by $N$ ``spins'' $\Psi_i$ where $i = 1, \dots, N.$  The spin operators
satisfy the  commutation relations
$$\{ S_{\pm}, S_0 \} = \pm S_{\pm}, \qquad \{ S_{+},S_{-} \} = \sinh 2 S_0$$
where the raising and lowering operators $S_{\pm} = S_1 \pm i S_2$ are defined as usual. Integrability of the spin chain can be shown starting from the auxiliary linear problem for the Lax matrix
$$L_i(\mu) \Psi_i(\mu) = \Psi_{i+1}(\mu)$$
with twisted boundary conditions implemented by the identification
$$\Psi_{i +N}(\mu) = -w \Psi_i(\mu).$$
The spectral curve of the spin chain is given by the determinant of the transfer matrix,
$$\det(T(\lambda) + w \,1) = 0,$$
where the transfer matrix is defined by the product of the two-by-two Lax matrices
$$T(\lambda) \equiv L_N(\lambda) \dots L_1(\lambda).$$

Impurities are added to the spin chain by performing a site-dependent shift of variables for the Lax matrices $L_j(\mu).$ $Y^{p,q}$ quiver gauge theories can be constructed by starting from $Y^{p,p}$ and adding $(p-q)$ impurities \cite{
Benvenuti:2004dy}. Amusingly, the spin chain and quiver impurities are precisely the same.  This reflects the well-known phenomena that the same integrable system can have different Lax representations.  Since the Lax matrices for all of the $Y^{p,q}$ quiver gauge theories are tridiagonal \cite{Franco:2005rj}, we can re-write the determinant of the Lax matrix in spin-chain form using the following identity
$$
\det
\begin{pmatrix}
a_1 & b_1 & & c_0 \\
c_1 & \ddots & \ddots & \\
 & \ddots & \ddots & b_{N-1} \\
 b_0 & & c_{N-1} & a_N
\end{pmatrix}
=
(-1)^{N-1}\left( \prod_j b_j + \prod_j c_j \right) + \Tr L_N L_{N-1} \dots L_1
$$
where
$$L_j = \begin{pmatrix}
a_j & - b_{j-1} c_{j-1} \\
1 & 0
\end{pmatrix} \, .
$$
Thus we can re-write the the spectral curve as
 $$\det(T(\lambda) + w 1) = w^2 + w \Tr T(\lambda) + \det T(\lambda).$$
 Under a suitable change of variables, it should be straightforward to show that this representation matches the XXZ form proposed in \cite{Gorsky:1997mw}. For the relativistic, periodic Toda chain this change of variables appears in \cite{Kuznetsov:1994ur}.

\section{Generating New Integrable Systems via Partial Resolution}

\label{section_partial_resolution}

In this section we explain how to determine the integrable system associated with a partial resolution, given the one for the parent theory. This is a useful way of obtaining new integrable systems from known ones. Starting from a relatively complicated example, partial resolution provides a practical way of deriving new integrable systems, much faster in practice that going through the process of expressing the new system in terms of loop variables. This method is so efficient that it is natural to expect that it has a counterpart in the integrable system literature. 

We will focus on minimal partial resolutions, which correspond to removing extremal perfect matchings (i.e those located at corners of the toric diagram) one at a time.\footnote{The implementation of more general partial resolutions using dimer models has been discussed in great detail in \cite{GarciaEtxebarria:2006aq}. We will restrict ourselves to minimal ones in this paper.} In addition, it might be necessary to simultaneously remove non-extremal perfect matchings. The result of this process is a new toric diagram in which the multiplicity of each new extremal perfect matching is one and some of the internal multiplicities might also change. From a quiver point of view, partial resolution corresponds to turning on non-zero vacuum expectation values (vevs) for  bifundamental fields $X_{ij}$ and then higgsing nodes $i$ and $j$.\footnote{Notice that, given a quiver, not all possible higgsings correspond to consistent partial resolutions.} This operation might result in mass terms for some fields, which can be integrated out using their equations of motion. From a dimer model perspective, partial resolution corresponds to removing from the tiling the edges associated to the fields with non-zero vevs. As a result, some of the adjacent faces in the tiling are merged into a single new face. 
The integration of massive fields maps to the removal of 2-valent vertices that might be generated in the process by condensing the nodes at the endpoints of the two edges terminating in them.  

The standard understanding of partial resolutions using dimer models is that all perfect matchings containing the edge associated with $X_{ij}$ are removed \cite{Franco:2005rj}. Following the connection between bifundamental fields and perfect matchings given in \cite{Franco:2005rj}, we see that all these perfect matchings (which are interpreted as GLSM fields) need to acquire a non-zero vev in order for the bifundamental to get one. 

In the integrable systems context, the fundamental objects are closed loops rather than individual perfect matchings. In fact, even the reference perfect matching might disappear when applying the discussion in the previous paragraph. It is straightforward to adapt the previous reasoning to loops: partial resolution removes all loops that contain an edge that gets a non-zero vev. 

\begin{figure}[h]
\begin{center}
\includegraphics[width=9cm]{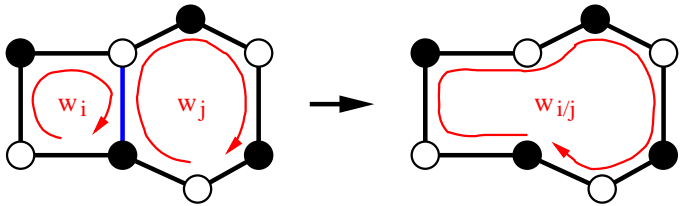}
\caption{Combination of gauge group cycles when higgsing by a vev associated to the blue edge.}
\label{dimer_higgsing}
\end{center}
\end{figure}

From a brane tiling perspective, the edge associated to $X_{ij}$ is deleted and faces $i$ and $j$ are combined into a single face. Consequently, we start from two cycles $w_i$ and $w_j$ and end with a combined cycle $w_{i/j}=w_i w_j$ as shown in \fref{dimer_higgsing}.

The following rules produce the integrable system for the partially resolved geometry:

\bigskip
{\bf 1)} Remove loops that contain an edge with a non-zero vev. \\

{\bf 2)} Re-express the surviving loops with the replacement $(w_i w_j)\to w_{i/j}$.
\bigskip

\noindent More practically, the loops that are removed in rule {\bf (1)} are those that cannot be re-written using rule {\bf (2)}. 

\begin{figure}[h]
\begin{center}
\includegraphics[width=9cm]{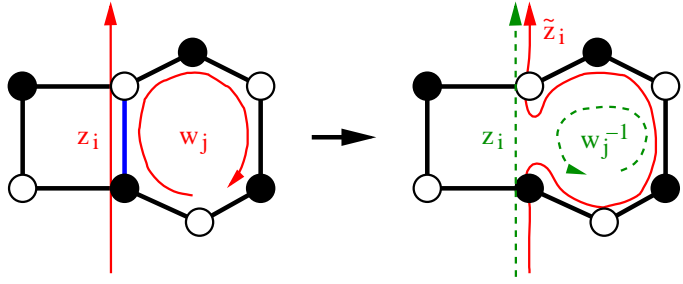}
\caption{Redefinition of one of the paths that winds around the $\mathbb{T}^2$, $z_i=\tilde{z}_i w_j$, after partial resolution by turning on a vev for the blue edge.}
\label{dimer_higgsing_z}
\end{center}
\end{figure}

In some cases, a $z_i$ path can involve an edge that is removed when higgsing, as shown in \fref{dimer_higgsing_z}. If so, the path can be redefined by using a $w_j$ to make the path wiggle appropriately, avoiding this edge. Equivalently, we could have chosen a different set of $z_i$ paths in the parent theory such that they do not involve higgsed edges.

\subsection{Examples: Partial Resolutions of $Y^{4,0}$}

We now illustrate our ideas in the explicit case of partial resolutions of $Y^{4,0}$. Partial resolution can either preserve or reduce the genus of the spectral curve (i.e. the number of Hamiltonians). The examples that follow exhibit the latter behavior.

In this section we depart from the notation for edges we used for general $Y^{p,0}$ in Section \ref{section_Kasteleyn_Toda}, which was specially devised for giving the Kasteleyn matrix a nice form. Here our emphasis is on higgsing, so we explicitly indicate the gauge groups under which bifundamentals are charged using subindices. $V_{ij}$ and $\tilde{V}_{ij}$ indicate vertical edges in the first and second columns of the brane tiling, respectively. Horizontal edges are denoted $H_{ij}$.

\begin{figure}[h]
\begin{center}
\includegraphics[width=3.5cm]{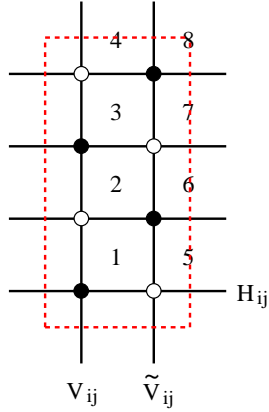}
\caption{Brane tiling for $Y^{4,0}$.}
\label{dimer_Y40}
\end{center}
\end{figure}

\begin{figure}[h]
\begin{center}
\includegraphics[width=9.5cm]{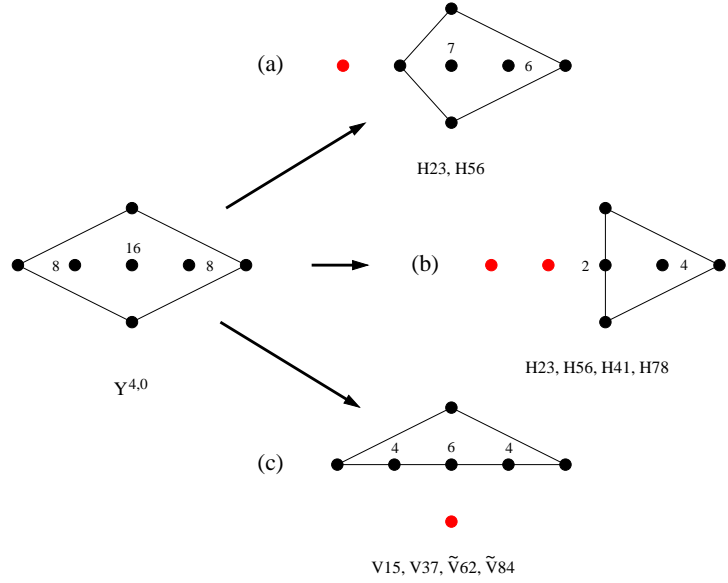}
\caption{Toric diagrams for $Y^{4,0}$ and various of its partial resolutions. We indicate the multiplicity associated with each point and the non-zero vevs that need to be turned on. Higgsing takes the number of Hamiltonians from 3 in the original theory to a) 2, b) 1 and c) 0.}
\label{toric_resolutions_Y40}
\end{center}
\end{figure}

\fref{toric_resolutions_Y40} shows the resolutions we will consider. The number of gauge groups in the associated quivers is given by twice the area of the toric diagrams. This implies that the number of bifundamental expectation values that need to be turned on is equal to twice the decrease in area of the toric diagram.

The starting point is the integrable system for $Y^{4,0}$. In the table below, $(n_1,n_2)$ gives the $(z_1^{n_1}z_2^{n_2})$ contribution. Every term in a given contribution arises from a loop in the tiling or equivalently from a perfect matching. Hamiltonians correspond to internal points and Casimirs are given by the ratio of consecutive points on the boundary. Instead of the basis of cycles used in Section \ref{section_Yp0}, here we use the $w_i$ (which correspond to gauge groups), $z_1$ and $z_2$ basis, since it makes higgsing more transparent.

\beq
\begin{array}{|c|c|}
\hline
\ \ \ (n_1,n_2) \ \ \ & \mbox{Loops} \\ \hline \hline 
(0,0) & 1 \\ \hline 
(-1,0) & \begin{array}{c} w_4+ w_4 w_8+ w_4 w_7 w_8+ w_3 w_4 w_7 w_8+ w_2 w_3 w_4 w_7 w_8 \\ +w_2 w_3 w_4 w_6 w_7 w_8
+w_2 w_3 w_4 w_5 w_6 w_7 w_8+1  \end{array} \\ \hline
(-2,0) & \begin{array}{c} w_1^{-1}w_5^{-1} w_4+ w_4 w_8 +  w_1^{-1}w_4 w_8+w_1^{-1} w_5^{-1} w_4 w_8 \\ +w_1^{-1} w_5^{-1} w_6^{-1} w_4 w_8+w_4 w_7 w_8
+w_1^{-1}w_4 w_7 w_8+w_1^{-1}w_5^{-1}w_4 w_7 w_8 \\ +w_3 w_4 w_7 w_8 
+w_1^{-1} w_3 w_4 w_7 w_8
+w_1^{-1} w_5^{-1}w_3 w_4 w_7 w_8+w_2 w_3 w_4 w_7 w_8\\ +w_1^{-1} w_5^{-1}+w_3 w_4^2 w_7 w_8+w_2 w_3 w_4^2 w_7 w_8+w_3 w_4^2 w_7 w_8^2 
\end{array} \\ \hline
(-3,0) & \begin{array}{c} w_1^{-1}w_5^{-1} w_4 w_8+  w_1^{-1}w_5^{-1}w_6^{-1} w_4 w_8+ w_1^{-1}w_5^{-1} w_4 w_7 w_8 + w_1^{-1}w_5^{-1} w_3 w_4 w_7 w_8 \\+ w_1^{-1}w_5^{-1} w_3 w_4^2 w_7 w_8+w_3 w_4^2 w_7 w_8^2+w_1^{-1} w_3 w_4^2 w_7 w_8^2 +w_1^{-1} w_5^{-1} w_3 w_4^2 w_7 w_8^2 \end{array} \\ \hline
(-4,0) & w_1^{-1} w_5^{-1} w_3 w_4^2 w_7 w_8^2 \\ \hline
(-2,1) &  w_1^{-1} w_4 w_7 w_8\\ \hline
(-2,-1) &  w_2 w_3 w_4^2 w_7 w_8 \\ \hline
\end{array}
\label{pms_Y40}
\eeq

\bigskip

\subsubsection*{Higgsing a}

\begin{figure}[h]
\begin{center}
\includegraphics[width=3cm]{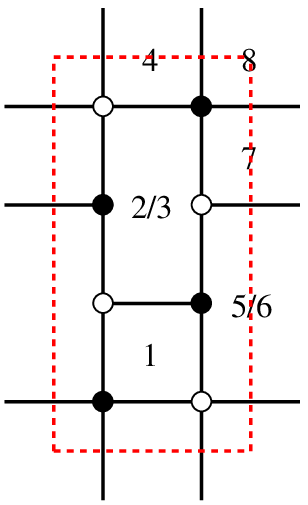}
\caption{Brane tiling for model (a) obtained by higgsing $Y^{4,0}$.}
\label{dimer_model_a}
\end{center}
\end{figure}

Model (a) in \fref{toric_resolutions_Y40} is obtained by giving non-zero vevs to both $H_{23}$ and $H_{56}$.  There are other choices of expectation values that lead to the same result. \fref{dimer_model_a} shows the corresponding brane tiling. 

The resulting integrable model is given by:

\beq
\begin{array}{|c|c|}
\hline
\ \ \ (n_1,n_2) \ \ \ & \mbox{Loops} \\ \hline \hline 
(0,0) & 1 \\ \hline 
(-1,0) & \begin{array}{c} w_4+ w_4 w_8+ w_4 w_7 w_8+ w_{2/3} w_4 w_7 w_8 \\ 
+w_{2/3} w_4 w_{5/6} w_7 w_8+1  \end{array} \\ \hline
(-2,0) & \begin{array}{c} w_4 w_8 +  w_1^{-1}w_4 w_8 +w_1^{-1} w_{5/6}^{-1} w_4 w_8+w_4 w_7 w_8 \\
+w_1^{-1}w_4 w_7 w_8   
+w_{2/3} w_4 w_7 w_8+w_{2/3} w_4^2 w_7 w_8 
\end{array} \\ \hline
(-3,0) & \begin{array}{c}  w_1^{-1}w_{5/6}^{-1} w_4 w_8 \end{array} \\ \hline
(-2,1) &  w_1^{-1} w_4 w_7 w_8\\ \hline
(-2,-1) &  w_{2/3} w_4^2 w_7 w_8 \\ \hline
\end{array}
\label{pms_A}
\eeq

\bigskip

\subsubsection*{Higgsing b}

\begin{figure}[h]
\begin{center}
\includegraphics[width=3cm]{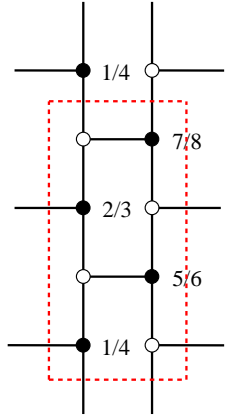}
\caption{Brane tiling for model (b) obtained by higgsing $Y^{4,0}$.}
\label{dimer_model_b}
\end{center}
\end{figure}

Model (b) corresponds to turning on vevs for $H_{23}$, $H_{56}$, $H_{78}$ and $H_{41}$. The resulting brane tiling is shown in \fref{dimer_model_b}. 

The edge associated to $H_{41}$ is contained in the original $z_1$ path, which thus needs to be redefined. We can consider a new path $\tilde{z}_1$ given by $z_1=\tilde{z}_1 w_1^{-1}$. The integrable system is summarized in \eref{pms_B}, where now $(n_1,n_2)$ corresponds to the $(\tilde{z}_1^{n_1}z_2^{n_2})$ contributions.

\beq
\begin{array}{|c|c|}
\hline
\ \ \ (n_1,n_2) \ \ \ & \mbox{Loops} \\ \hline \hline 
(0,0) & 1 \\ \hline 
(-1,0) & \begin{array}{c} w_{4/1}+ w_{4/1} w_{7/8}+ w_{2/3} w_{4/1} w_{7/8} 
+w_{2/3} w_{4/1} w_{5/6} w_{7/8}  \end{array} \\ \hline
(-2,0) & \begin{array}{c} w_{4/1} w_{7/8} + w_{2/3} w_{4/1}^2 w_{7/8} 
\end{array} \\ \hline
(-2,1) &  w_{4/1} w_{7/8} \\ \hline
(-2,-1) &  w_{2/3} w_{4/1}^2 w_{7/8} \\ \hline
\end{array}
\label{pms_B}
\eeq

\bigskip

\subsubsection*{Higgsing c}

\begin{figure}[h]
\begin{center}
\includegraphics[width=3cm]{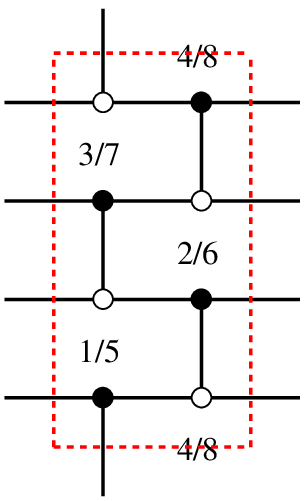}
\caption{Brane tiling for model (c) obtained by higgsing $Y^{4,0}$.}
\label{dimer_model_c}
\end{center}
\end{figure}

Finally, model (c) follows from turning on vevs for $V_{15}$, $V_{37}$, $\tilde{V}_{62}$ and $\tilde{V}_{84}$. The corresponding brane tiling is shown in \fref{dimer_model_c}.

This time both $V_{15}$ and $V_{37}$ overlap with the original $z_2$ path, which can be redefined according to $z_2=\tilde{z}_2 w_1 w_3$. Denoting $(n_1,n_2)$ the $(z_1^{n_1}\tilde{z}_2^{n_2})$ contribution, \eref{pms_C} summarizes the resulting integrable system.

\beq
\begin{array}{|c|c|}
\hline
\ \ \ (n_1,n_2) \ \ \ & \mbox{Loops} \\ \hline \hline 
(0,0) & 1 \\ \hline 
(-1,0) & \begin{array}{c} w_{8/4}+ w_{3/7} w_{8/4}+w_{6/2} w_{3/7} w_{8/4}+1  \end{array} \\ \hline
(-2,0) & \begin{array}{c}  w_{8/4}+ w_{1/5}^{-1} w_{8/4} +w_{3/7} w_{8/4} 
+w_{1/5}^{-1} w_{3/7} w_{8/4}+w_{1/5}^{-1}+w_{3/7} w_{8/4}^2 
\end{array} \\ \hline
(-3,0) & \begin{array}{c} w_{1/5}^{-1} w_{8/4}+ w_{1/5}^{-1} w_{3/7} w_{8/4} +w_{3/7} w_{8/4}^2 +w_{1/5}^{-1} w_{3/7} w_{8/4}^2 \end{array} \\ \hline
(-4,0) & w_{1/5}^{-1} w_{3/7} w_{8/4}^2 \\ \hline
(-2,1) &  w_{3/7} w_{8/4} \\ \hline
\end{array}
\label{pms_C}
\eeq

From an integrability point of view, this model is trivial, i.e. it only consists of Casimirs. From a quiver perspective, the reason for this is that, as one can deduce from \fref{dimer_model_c}, the associated quiver is fully non-chiral. This implies that all commutators vanish.

\section{Conclusions and Outlook}

We have investigated various applications of the correspondence between dimer models and integrable system introduced by Goncharov and Kenyon in \cite{GK}. We used it to explicitly construct relativistic generalizations of the periodic Toda chain associated to $Y^{p,0}$ and $Y^{p,p}$  geometries. In these models, the calculation of commuting Hamiltonians reduces to the combinatorics of non-intersecting paths on the brane tiling. We investigated the connection between the Kasteleyn matrix and the Lax operator, the non-relativistic limit of the integrable systems, additional relativistic versions of the periodic Toda chain based dimer models for general $Y^{p,q}$ geometries, and the identification of quiver impurities and spin chain impurities. Finally, we introduced a method for generating new integrable systems based on higgsing. We can envision, and are currently pursuing, multiple directions in which the correspondence between dimer models and integrable systems can be exploited. We discuss some of them below.

We have explained how higgsing is an efficient tool for generating new integrable systems. The characteristic polynomials for the dimer models associated to $\mathbb{Z}_n\times \mathbb{Z}_m$ orbifolds of arbitrary geometries, which correspond to $n\times m$ arrays of copies of the original unit cell, can be determined using simple formulas \cite{Kenyon:2003uj}. These expressions have been used for calculating the multiplicity of perfect matchings associated to points in the toric diagrams of orbifolds \cite{Hanany:2005ve}. It would be interesting to investigate whether there are analogous expressions for the integrable systems associated to orbifolds starting from the integrable system for the unorbifolded geometry. New integrable systems could then be generated by higgsing, using the orbifold theories as starting points.   

It would be interesting to study the continuous limit of integrable systems arising from dimer models. By this, we not only mean the infinite length limit of integrable chains but also the (1+1)-dimensional field theory limit of fixed length systems. It is reasonable to conjecture that field theories such as $A_n$ Toda field theories can be constructed in this way. Furthermore, it is natural to expect that dimer models are useful for classifying (0+1)-dimensional, integrability-preserving defects and interfaces that can be added to such field theories.

The work of \cite{Dimofte:2011gm} investigated the quantization of Riemann surfaces defined by the vanishing of the A-polynomials of three-manifolds $M$ that are the complement of (thickened) knots or links. In the case of knots, the boundary of $M$ is a 2-torus. The decomposition of $M$ into glued tetrahedra with truncated vertices gives rise to a triangulation of the 2-torus, the developing map, whose dual is reminiscent of a dimer model. On the other hand, \cite{GK} and \cite{Franco:2011sz} discussed the relation between dimer models and quantum Teichm\"uller space. It is then natural to ask whether the existing similarities indicate the existence of a true connection that associates dimer models to three-manifolds that are knot complements. 

One clear direction for further research is to explore the connection between quantum integrable systems and gauge theories proposed by Nekrasov and Shatashvili \cite{Nekrasov:2009rc}.  By considering the 5d gauge theory on an $\Omega$-background with $\epsilon_1 = \epsilon$ and $\epsilon_2 = 0,$ the classical integrable systems we have investigated become quantized.  We now review the interpretation of the quantization in terms of a B-brane on the spectral curve using the refined topological string \cite{Aganagic:2011mi}.  The mirror Calabi-Yau geometry takes the form
$$uv + H(x,p) = 0.$$  When the $\Omega$-background is turned on, the classical spectral curve, $\Sigma$ defined by $H(x,p) = 0$, is promoted to a quantum spectral curve $$\hat{H}(x,p) \Psi(x) = 0$$
where $\Psi(x)$ is the wave-function for a B-brane. The quantum spectral curve is the Baxter equation for the relativistic Toda chain \cite{Kuznetsov:1994ur}.  The coefficients of $H(x,p)$ are the energy eigenvalues of the Hamiltonians (and Casimirs) of the quantum integrable system.  
In the relativistic case, the differential operators $e^{\hat{p}}$ become shift operators, acting on $\Psi(x)$ by $$e^{\hat{p}} \Psi(x) \rightarrow \Psi(x + \hbar).$$ 
Solutions to this differential equation were obtained from the refined topological string partition function in \cite{Aganagic:2011mi}. Thus the refined topological string should provide a way to solve the Baxter equation for quantum integrable systems. We plan to elucidate this connection in future work.

A graphical representation for the finite non-periodic Toda lattice, similar to dimer models, was developed \cite{Gekht10} using the Poisson geometry of planar directed networks in an annulus \cite{Gekht09}.  The geometry of planar directed networks was developed in order to study the totally nonnegative Grassmannian \cite{post06}.  Our work suggests an intriguing connection between the geometry of brane tilings and the totally nonnegative part of the double loop Grassmannian.

In general, gauge theories arising from brane tilings have sequences of periodic Seiberg dualities known as duality cascades \cite{Klebanov:2000hb}-\cite{Franco:2005fd}. According to \cite{GK}, Seiberg dualities correspond to canonical transformations of the integrable system.  In terms of the new variables the Hamiltonians will typically take a different functional form.  However, since cascades are periodic, we are interested in special canonical transformations that preserve the functional form of the commuting Hamiltonians.  Such canonical transformations are known as auto-B\"acklund-Darboux transformations.  For example, the transformation \cite{MR1755477} 
$$\tilde{c}_i =c_i \frac{d_i + c_{i-1}}{d_{i+1}+c_i}, \qquad \tilde{d}_i = d_{i+1} \frac{d_i + c_{i-1}}{d_{i+1}+c_i}$$
is an auto-B\"acklund-Darboux transformation of the relativistic Toda chain because it is canonical and the new Hamiltonian
$$\tilde{H} = \sum_i \left( \tilde{c}_i + \tilde{d}_i \right)$$
takes the same functional form as the original Hamiltonian. The theory of auto-B\"acklund-Darboux transformations is closely related to the theory of separation of variables and discrete-time integrable systems \cite{MR1160332,Kuzn98}.  Thus we expect a fruitful interplay between duality cascades and integrable systems.

\section*{Acknowledgments}

We would like to thank Y. H. He and D. Morrison for useful discussions. We also thank D. Morrison for reading the manuscript. We are particularly grateful to M. Aganagic for early collaboration and insightful discussions. S. F. would like to thank to A. Goncharov and R. Kenyon for sharing a draft of \cite{GK} before its publication. The work of R. E. is supported in part by the National Science Foundation under
grant DMS-1007414 and by the World Premier International Research
Center Initiative (WPI Initiative), MEXT, Japan. S. F. is supported by the National Science Foundation under Grant No. PHY05-51164. The work of K. S. is supported by the Berkeley Center for Theoretical Physics.

\bigskip


\end{document}